\documentclass[twocolumn,twoside,slac_two]{revtex4}
\usepackage{graphicx}
\usepackage{fancyhdr}
\newcommand{\be}{\begin{equation}}
\newcommand{\ee}{\end{equation}}
\newcommand{\bea}{\begin{eqnarray}}
\newcommand{\eea}{\end{eqnarray}}
\newcommand{\bwt}{\begin{widetext}}
\newcommand{\ewt}{\end{widetext}}

\begin{document}

\title{Nonlinearly realized electroweak symmetry and supersymmetric multi-Higgs doublet models}

\author{S.T. Love}
\affiliation{Department of Physics,
 Purdue University,  
 West Lafayette, IN 47907-2036, U.S.A.}

\begin{abstract}
The  minimal supersymmetric standard model (MSSM) is extended by including an additional pair of constrained Higgs doublet superfields through which the electroweak symmetry is nonlinearly realized.  The superpotential couplings of this constrained Higgs doublet pair to the MSSM Higgs doublet pair catalyze the later nontrivial vacuum expectation values.  The  mass spectrum of the Higgs scalars and Higgsino-gaugino sector is presented for several choices of supersymmetry (SUSY) breaking and Higgs superpotential mass parameters.  The effect of the the additional  constrained fields and multiple vacuum expectation values  on the lowest mass neutral Higgs scalar production and decay is presented.
\end{abstract}

\maketitle 

\thispagestyle{fancy}
As experimental results continue to squeeze the MSSM, particularly through the increasing lower bound on the Higgs boson mass, it proves worthwhile to focus increasing attention on SUSY models beyond the MSSM. Indeed, there are a plethora of such models already introduced, with the NMSSM\cite{NMSSM} garnering the most attention. Here we focus on another model (actually a class of models)\cite{CLtV} characterized by having the dynamics responsible for the origin of the electroweak symmetry breaking separate from the source of the quark and lepton masses.  The electroweak symmetry is assumed to be broken by some unspecified, presumably strong, supersymmetric dynamics which further respects the custodial $SU(2)_V$ global symmetry. The low energy effects of the electroweak symmetry breakdown are manifested through an additional pair of constrained Higgs doublet superfields which realize the electroweak symmetry nonlinearly. On the other hand, the quark and lepton superfields are Yukawa coupled only to the MSSM Higgs doublet superfields $H_u, H_d$.
\begin{figure*}
\begin{center}
$\begin{array}{cc}
\includegraphics[scale=0.75]{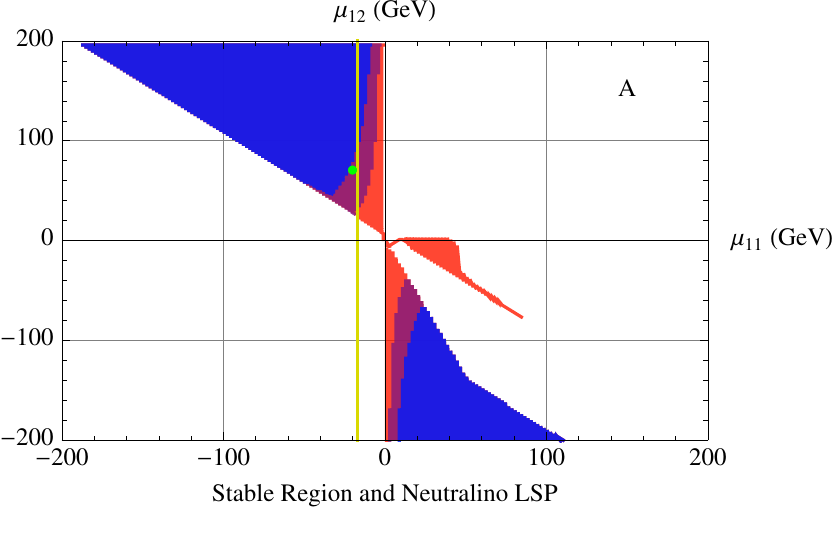} &
\includegraphics[scale=0.75]{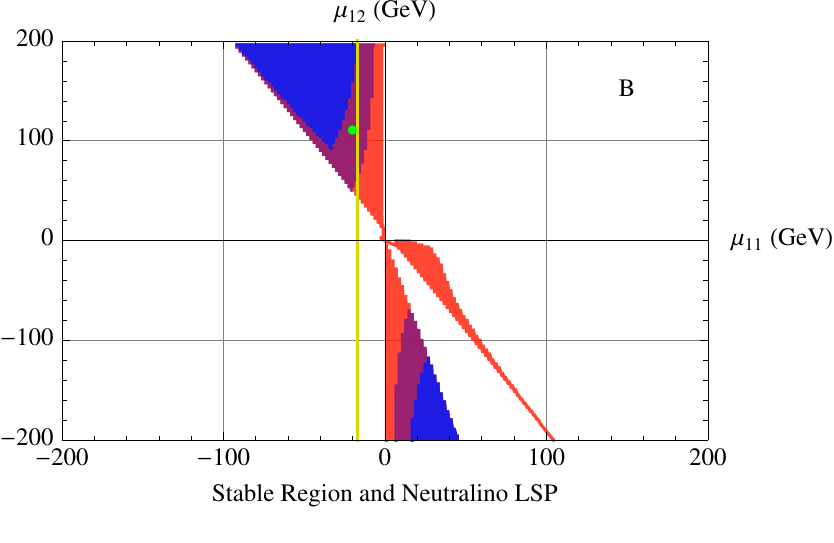} \\
\includegraphics[scale=0.75]{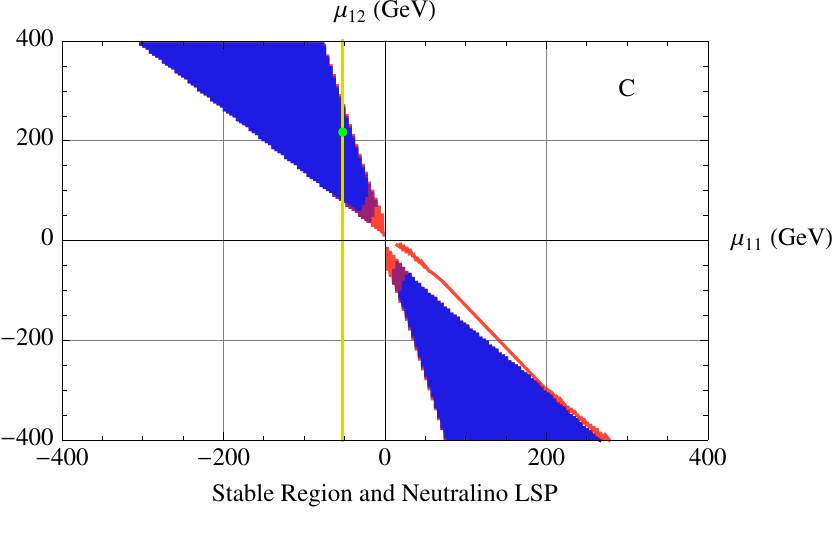} &
\includegraphics[scale=0.75]{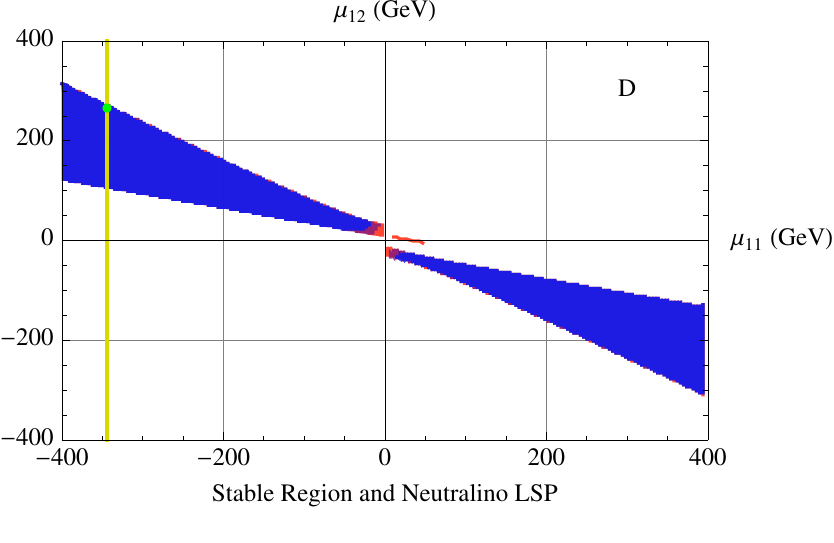} 
\end{array}$
\caption{Stability of the potential against $D$-flat direction runaway field values is determined in the $\mu_{11}$-$\mu_{12}$ parameter plane.  Each region of SUSY breaking parameter $b=-4,000,~4,000,~12,000$ ${\rm GeV^2}$ is depicted by the overlapping orange, violet, blue regions, respectively. Stability region A has $\tan{\beta}=1$, $\tan{\theta}=1$, region B has $\tan{\beta}=1$, $\tan{\theta}=2$, region C has $\tan{\beta}=2$, $\tan{\theta}=2$, and region D has $\tan{\beta}=10$, $\tan{\theta}=2$. The requirement that a neutralino is the LSP further delineates the stability regions. The green dots indicate the points in parameters space associated with the detailed mass spectrum in Fig. \ref{MassSpectrum}. The yellow lines indicate the value of $\mu_{11}$ along which the parameter $\mu_{12}$ is scanned in  the subsequent mass spectrum plots. For each plot the value of the gaugino SUSY breaking masses are $M_1 =200$ GeV and $M_2=800$ GeV. }
\label{StabilityandLSPRegion}
\end{center}
\end{figure*}
The additional pair of constrained doublet chiral superfields denoted $H_u^\prime, H_d^\prime$ have the form
$H_u^\prime = \left[\begin{array}{rr}
H_u^{+\prime}\\
H_u^{0\prime}
\end{array}\right]=\left[\begin{array}{rr}
i\Pi^+\\
\Sigma -i\Pi^0
\end{array}\right], 
H_d^\prime = \left[\begin{array}{rr}
H_d^{0\prime}\\
H_d^{-\prime}
\end{array}\right]=\left[\begin{array}{rr}
\Sigma+i\Pi^0\\
i\Pi^-
\end{array}\right]$, 
with the nontrivial vacuum expectation values $<0|H_u^{0\prime}|0> =
\frac{v^\prime }{\sqrt{2}}= <0|H_d^{0\prime}|0>$. 
The equality of the primed VEVs is a consequence of the assumed $SU(2)_V$ custodial symmetry of the electroweak breaking dynamics. Implementing the superfield constraint, $H_d^\prime \epsilon H_u^\prime = 
\frac{v^{\prime 2}}{2}$, as $\Sigma = \sqrt{\frac{v^{\prime 2}}{2}-\vec{\Pi} \cdot \vec{\Pi}~}$, 
 allows the $\Sigma$ superfield to be eliminated in favor of the $\vec{\Pi}$ superfields and breaks the electroweak symmetry. Note that, while in the MSSM, the electroweak symmetry breakdown is tied to the SUSY breaking so that without SUSY breaking there is no electroweak breaking,  this multi-doublet model can be realized in the broken electroweak symmetry phase even if SUSY remains unbroken. 

The Higgs doublet portion of the superpotential includes the mixing terms among the constrained and MSSM Higgs multiplets as well as the MSSM $\mu_{11}$-term so that 
$W=\mu_{11} H_u \epsilon H_d +\mu_{12} H_u \epsilon H_d^\prime + \mu_{21} H_u^\prime \epsilon H_d$. The VEVs of the MSSM Higgs doublets $<0|H_u^0|0> =
\frac{v_u }{\sqrt{2}} ; <0|H_d|0> =\frac{v_d}{\sqrt{2}}$, are catalyzed through their bilinear coupling to the  constrained doublets. Note that nontrivial values for $v_u, v_d$ require both nonzero $v^\prime$ and  nonvanishing mixing parameters $\mu_{12}, \mu_{21}$. The quarks and leptons  acquire masses through their couplings to the MSSM Higgs doublets. 
Finally, the soft SUSY breaking terms for the gauginos, $\lambda^i, \lambda$,  and component field MSSM Higgs doublets are given by 
${\cal L}_{\rm \rlap{/}{S}} = \frac{1}{2} M_1 \left( \lambda\lambda + \bar{\lambda}\bar{\lambda}\right) + \frac{1}{2} M_2 \left( \lambda^i\lambda^i + \bar{\lambda^i}\bar{\lambda^i}\right)  -m_u^2 H_u^\dagger H_u - m_d^2 H_d^\dagger H_d -\mu_{11} B H_u \epsilon H_d - \mu_{11} B H_u^\dagger \epsilon H_d^\dagger$. 

\begin{figure*}
\begin{center}
$\begin{array}{cc}
\includegraphics[scale=0.50]{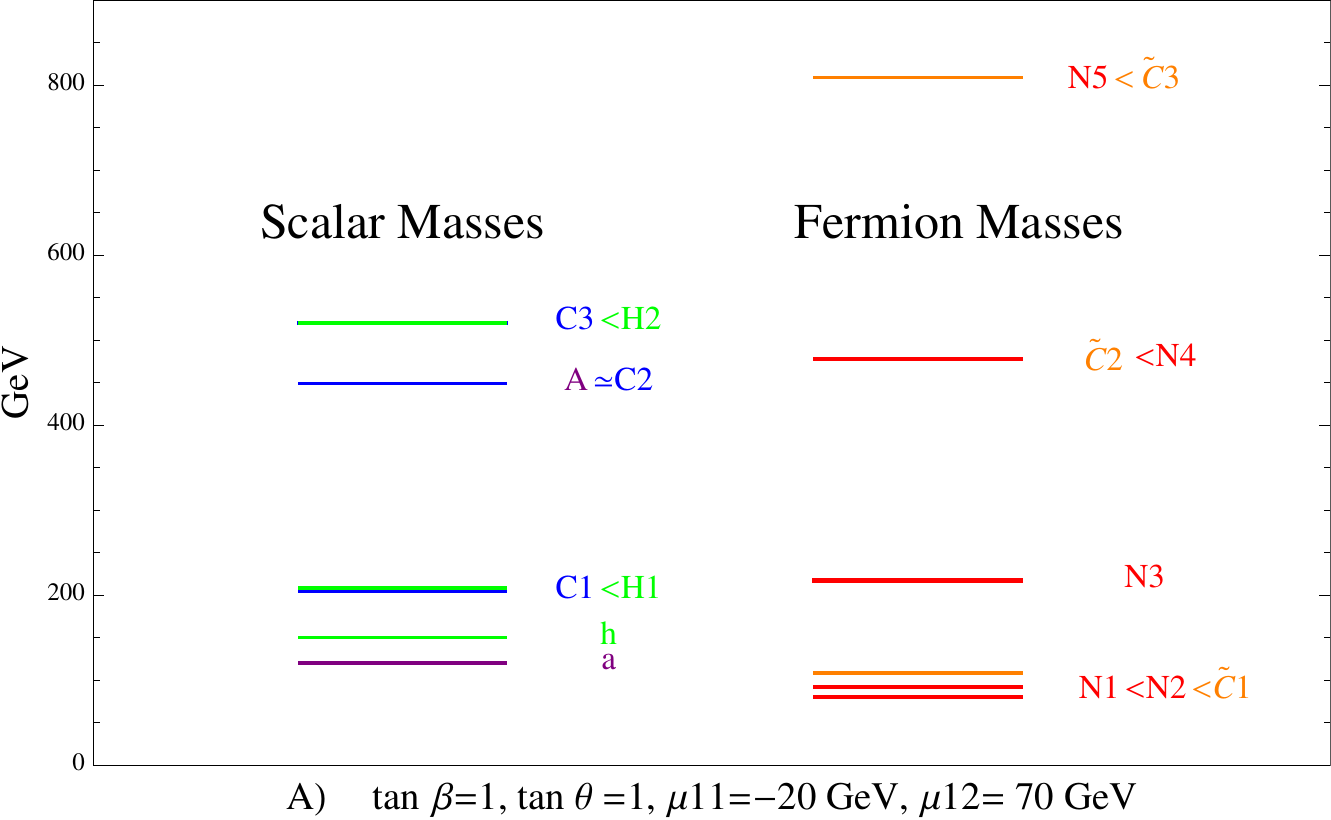} &
\includegraphics[scale=0.50]{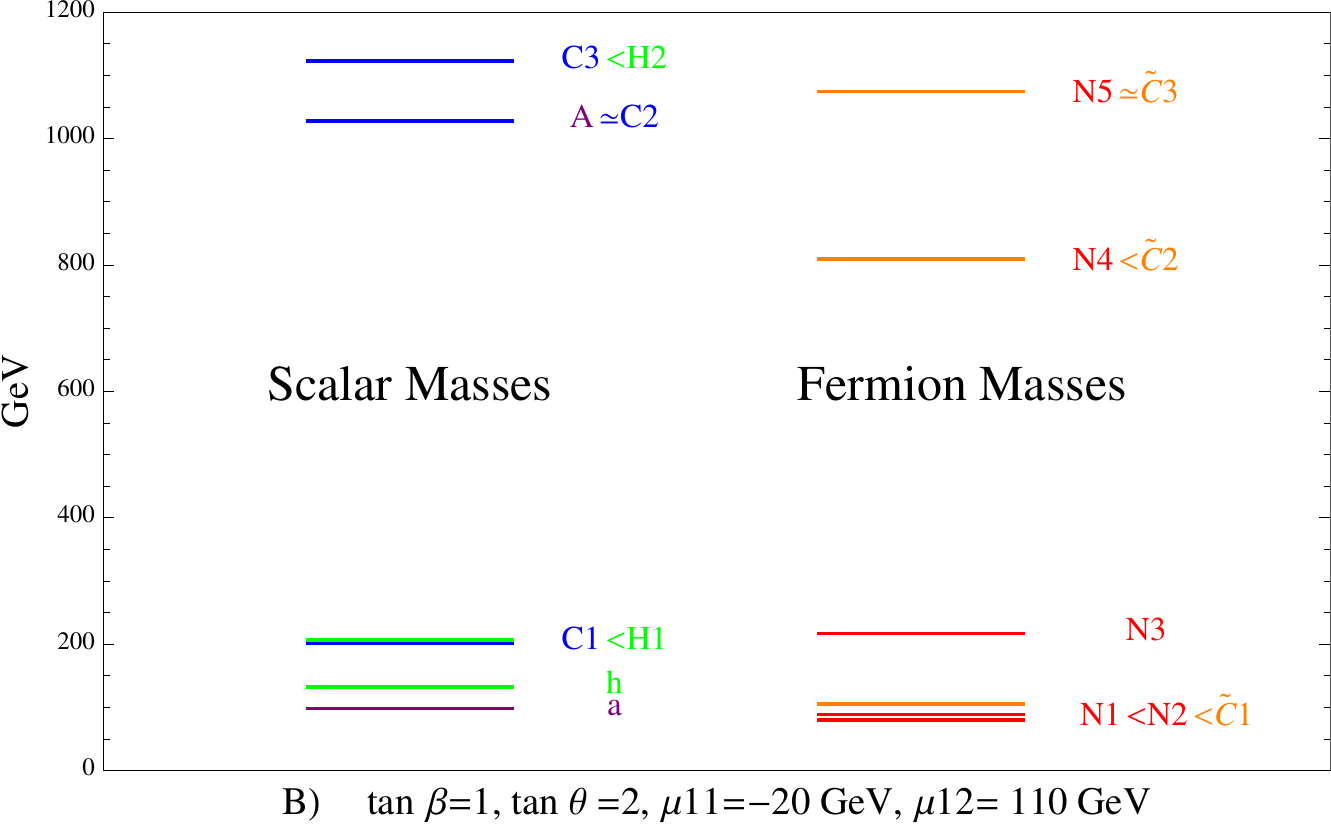} \\
\includegraphics[scale=0.50]{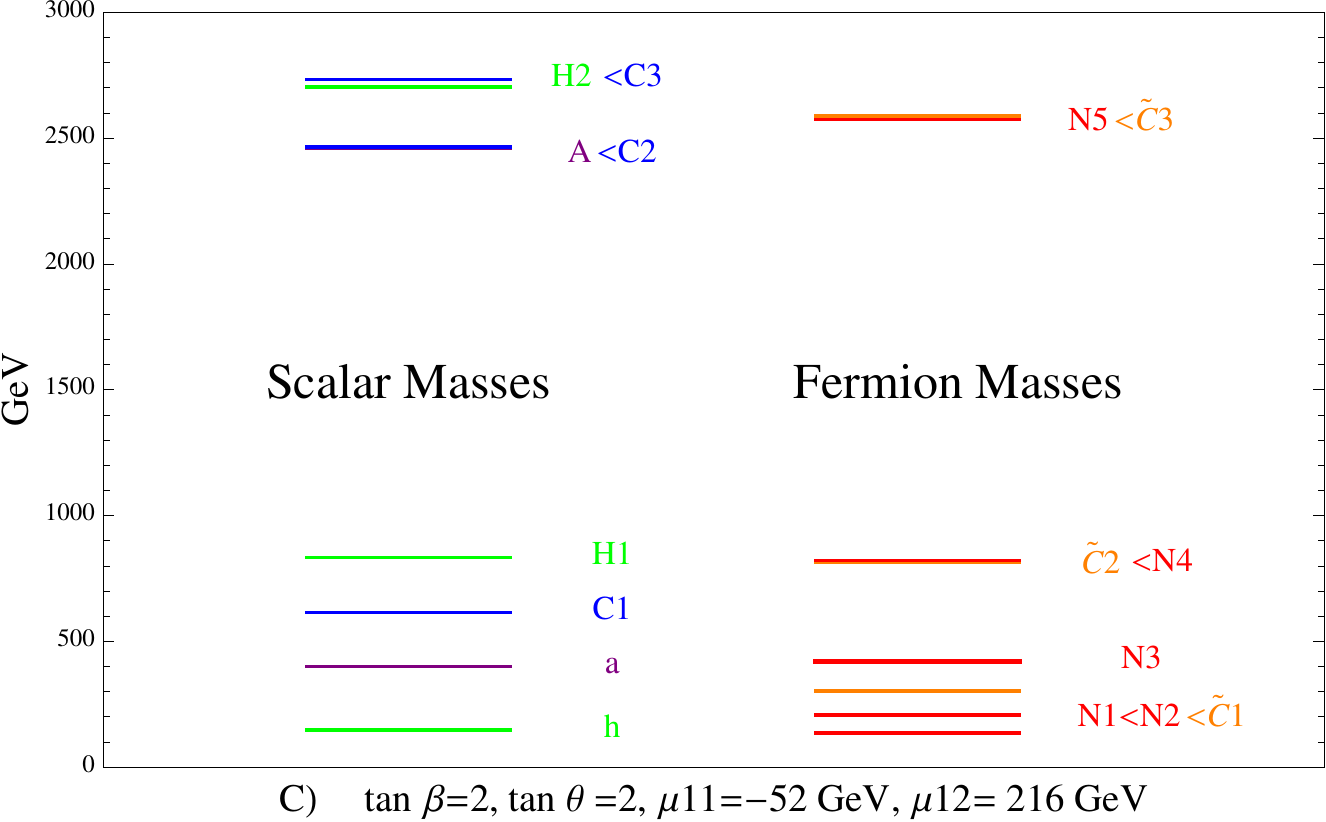} &
\includegraphics[scale=0.50]{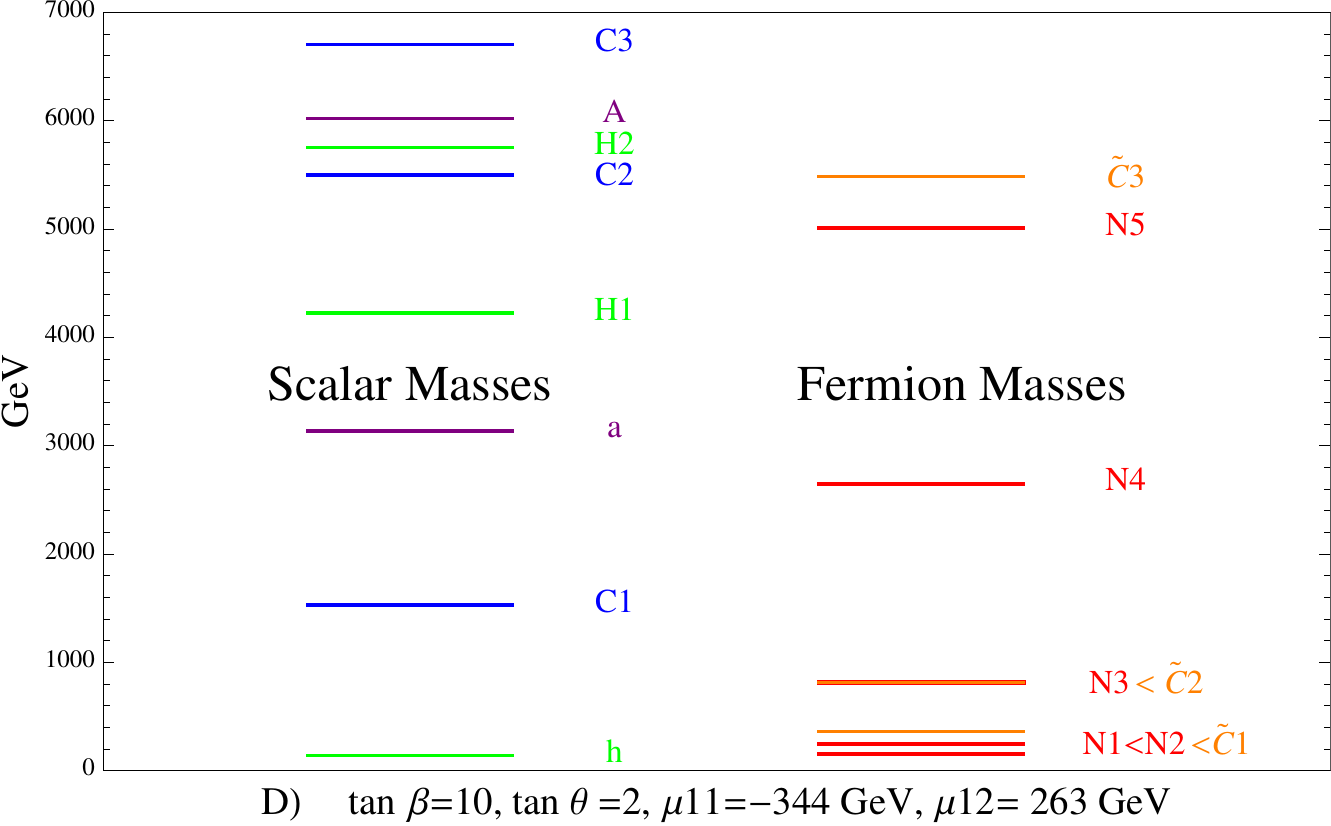} 
\end{array}$
\caption{The Higgs (pseudo-) scalars and gaugino-Higgsino mass spectrum for a point in the LSP-stability regions indicated by the green dot in 
Fig.~\ref{StabilityandLSPRegion}.  The gaugino soft SUSY breaking masses are $M_1 =200$ GeV and $M_2=800$ GeV, and $b=4,000 $ ${\rm GeV}^2$ for all regions.}
\label{MassSpectrum}
\end{center}
\end{figure*}
\begin{figure*}
\begin{center}
$\begin{array}{cc}
\includegraphics[scale=0.75]{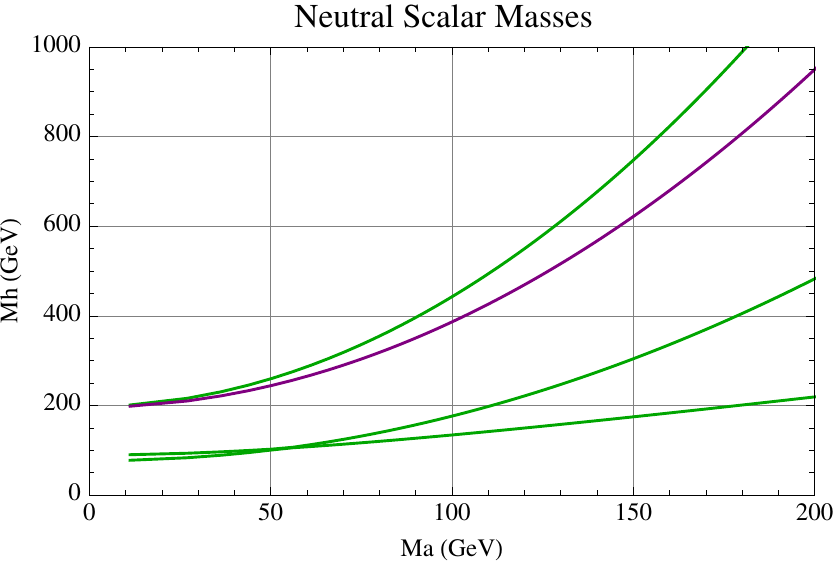} &
\includegraphics[scale=0.75]{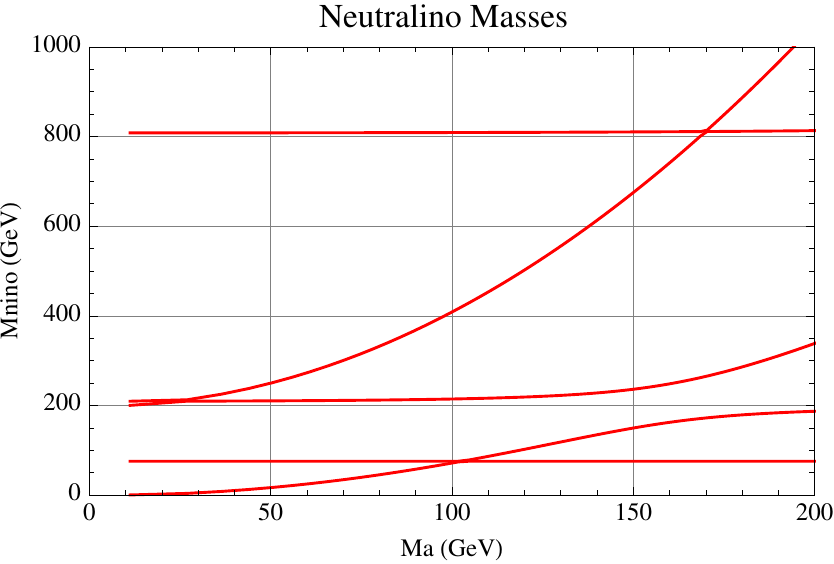} \\
\includegraphics[scale=0.75]{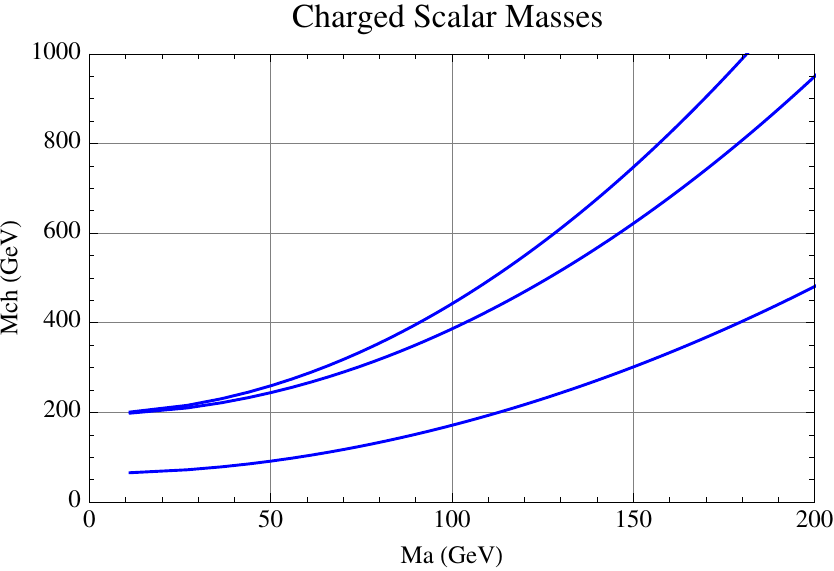} &
\includegraphics[scale=0.75]{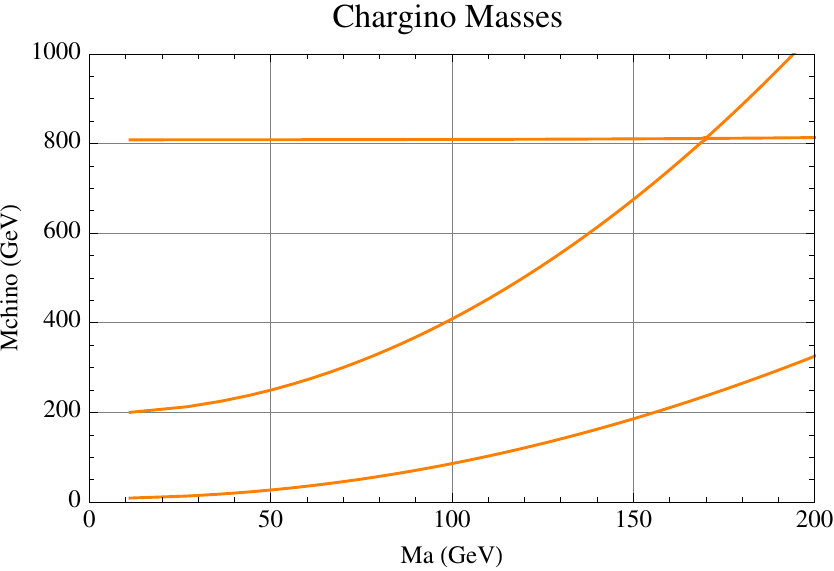} 
\end{array}$
\caption{Masses  as a function of the lightest pseudoscalar mass $m_a$ for a $\mu_{12}$ scan along the yellow line across region A in Fig. \ref{StabilityandLSPRegion}.  The parameters for the plots are $\tan{\beta}=1$, $\tan{\theta}=1$, $b=4,000$ ${\rm GeV^2}$ and $\mu_{11} =-12$ GeV. In the top left panel green curves correspond to scalar $h, H1, H2$ masses, while the purple curve corresponds to the  pseudoscalar 
$A$ mass. In the bottom left panel, the blue curves correspond to the charged Higgs C1, C2, C3 masses. In the top right panel, the red curves correspond to the neutralino $N1-N5$ masses, while the orange curves in the lower right panel correspond to the chargino $\tilde{C1}, \tilde{C2}, \tilde{C3}$ masses.}
\label{MassesTanBeta1}
\end{center}
\end{figure*}

\begin{figure*}
\begin{center}
$\begin{array}{cc}
\includegraphics[scale=0.75]{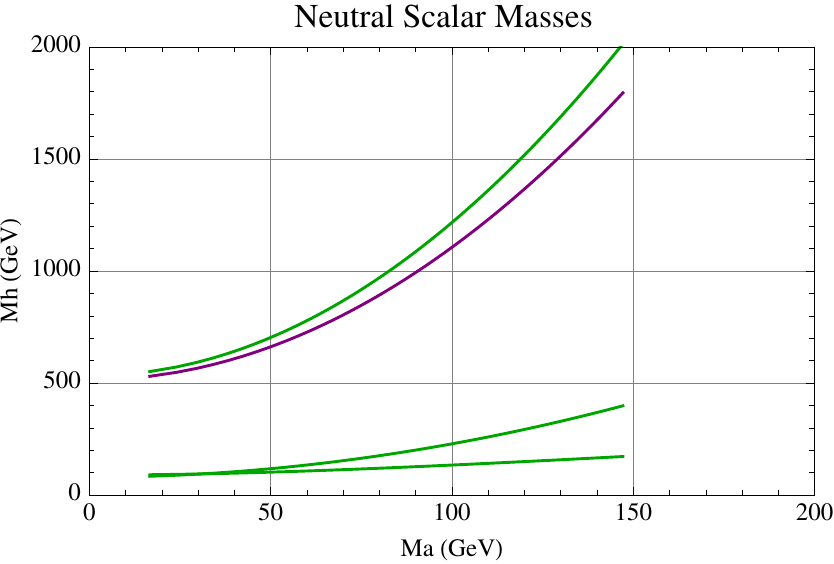} &
\includegraphics[scale=0.75]{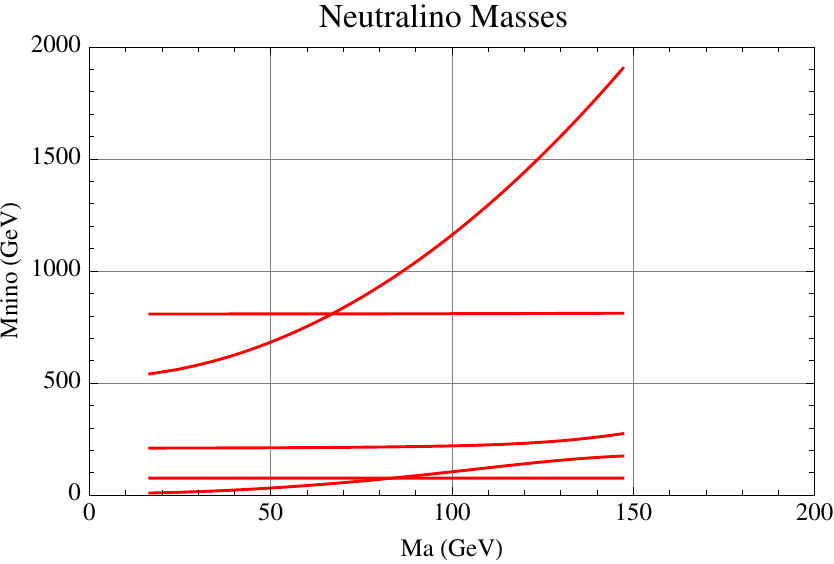} \\
\includegraphics[scale=0.75]{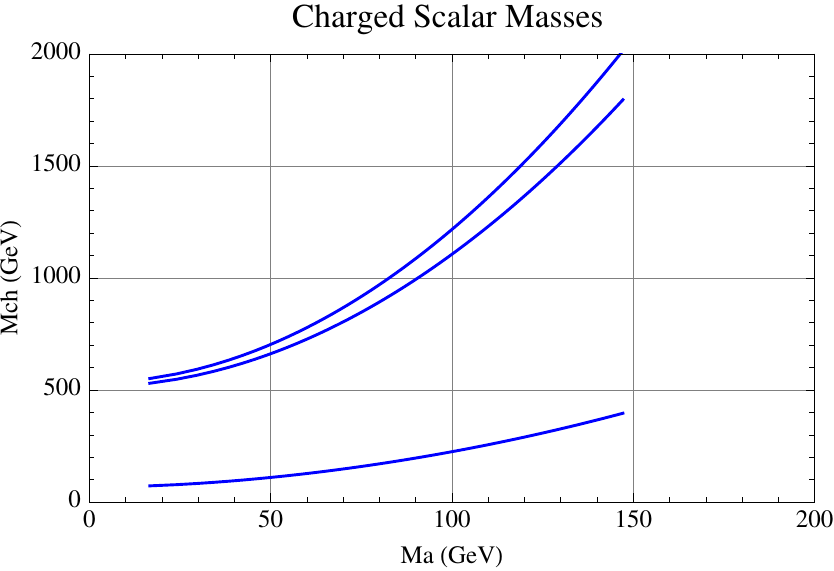} &
\includegraphics[scale=0.75]{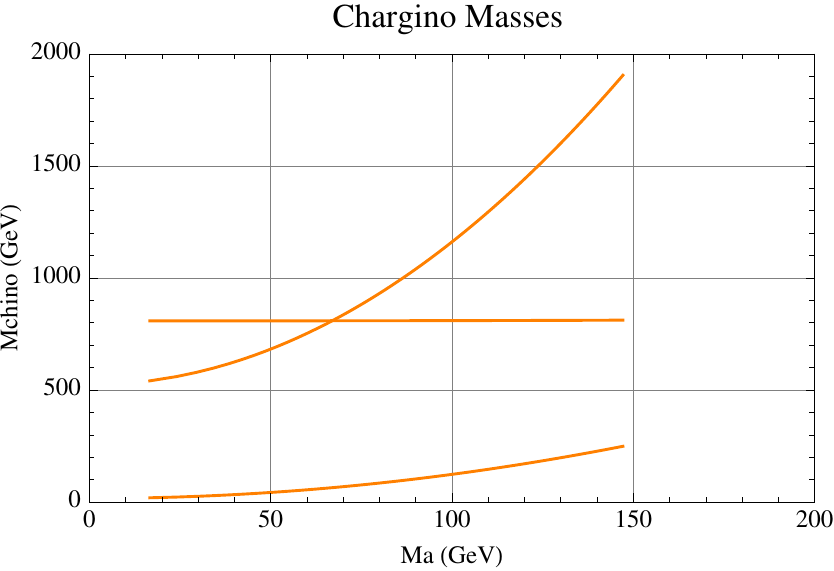} 
\end{array}$
\caption{Masses as a function of the lightest pseudoscalar mass $m_a$  for a $\mu_{12}$ scan along the yellow line across region B in Fig. \ref{StabilityandLSPRegion}.  The parameters for the plots are $\tan{\beta}=1$, $\tan{\theta}=2$, $b=4,000$ ${\rm GeV^2}$ and $\mu_{11} =-16$ GeV. The  curves correspond to the various particles  just as described in the caption to Fig. \ref{MassesTanBeta1}. }
\label{MassesTanBeta2}
\end{center}
\end{figure*}

\begin{figure*}
\begin{center}
$\begin{array}{cc}
\includegraphics[scale=0.75]{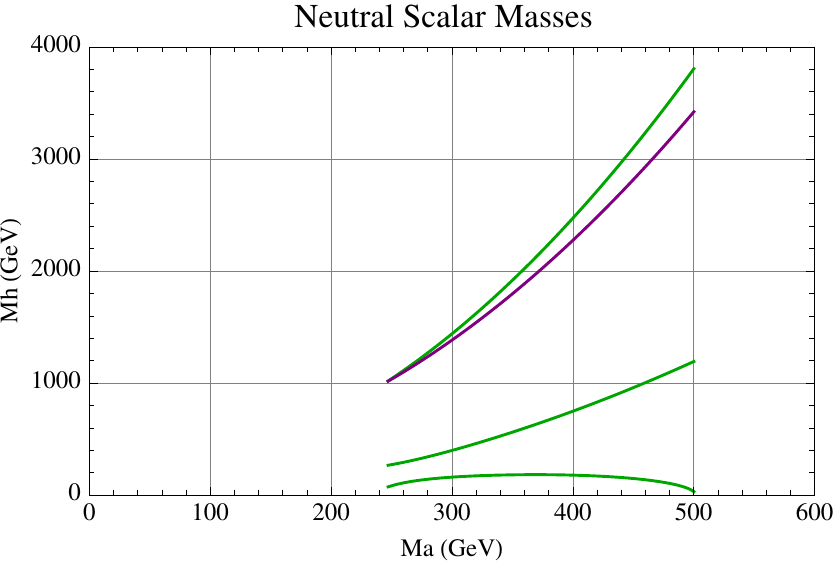} &
\includegraphics[scale=0.75]{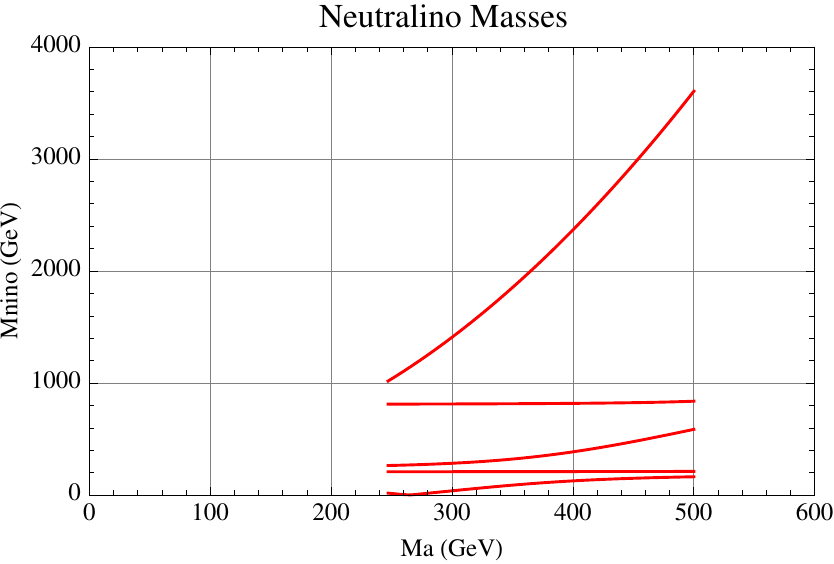} \\
\includegraphics[scale=0.75]{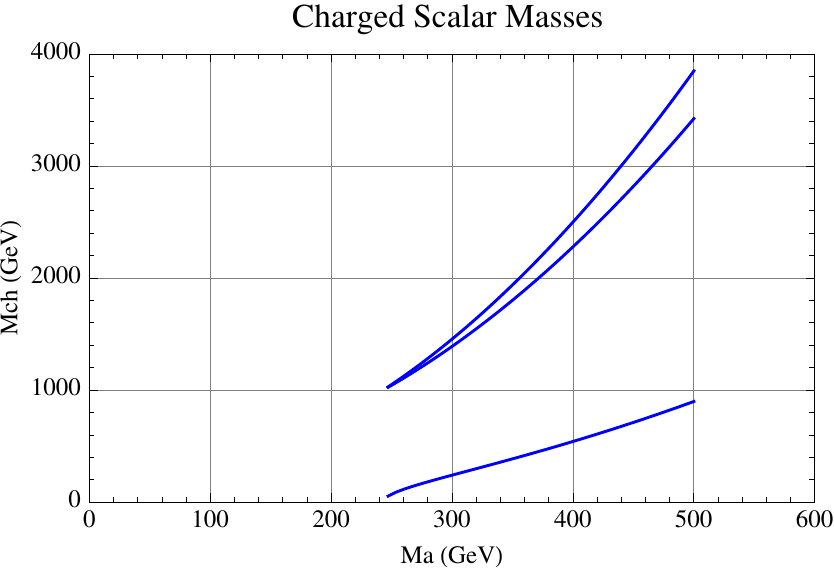} &
\includegraphics[scale=0.75]{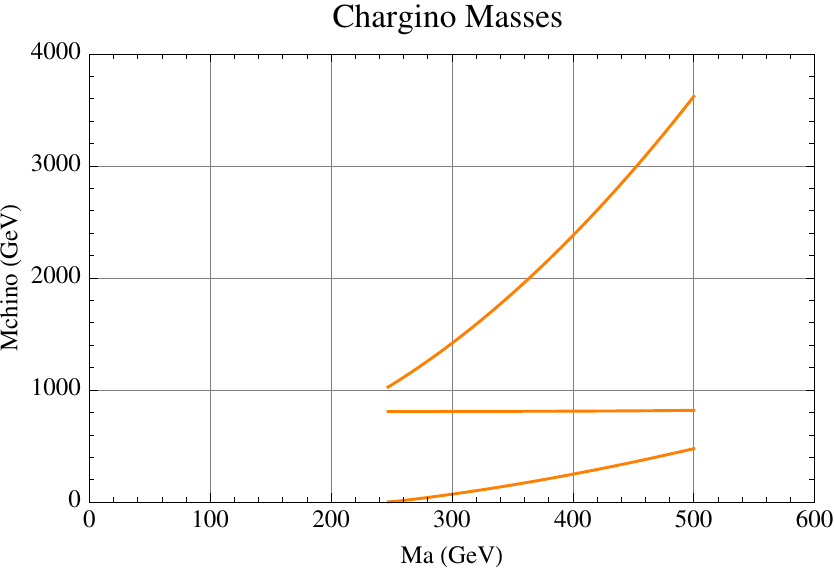} 
\end{array}$
\caption{Masses as a function of the lightest pseudoscalar mass $m_a$ for a $\mu_{12}$ scan along the yellow line across region C in Fig. \ref{StabilityandLSPRegion} .  The parameters for the plots are $\tan{\beta}=2$, $\tan{\theta}=2$, $b=4,000$ ${\rm GeV^2}$ and $\mu_{11} =-52$ GeV. The  curves correspond to the various particles  just as described in the caption to Fig. \ref{MassesTanBeta1}.}
\label{MassesTanBeta10}
\end{center}
\end{figure*}

\begin{figure*}
\begin{center}
$\begin{array}{cc}
\includegraphics[scale=0.75]{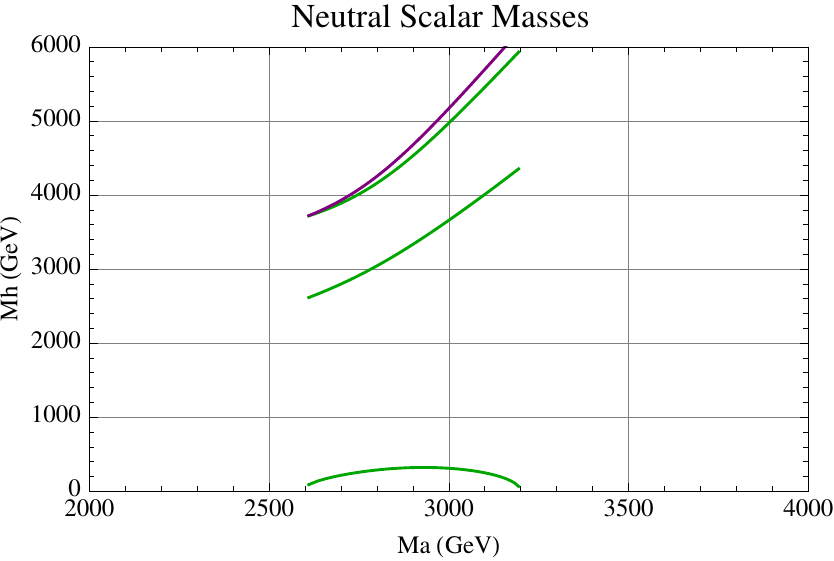} &
\includegraphics[scale=0.75]{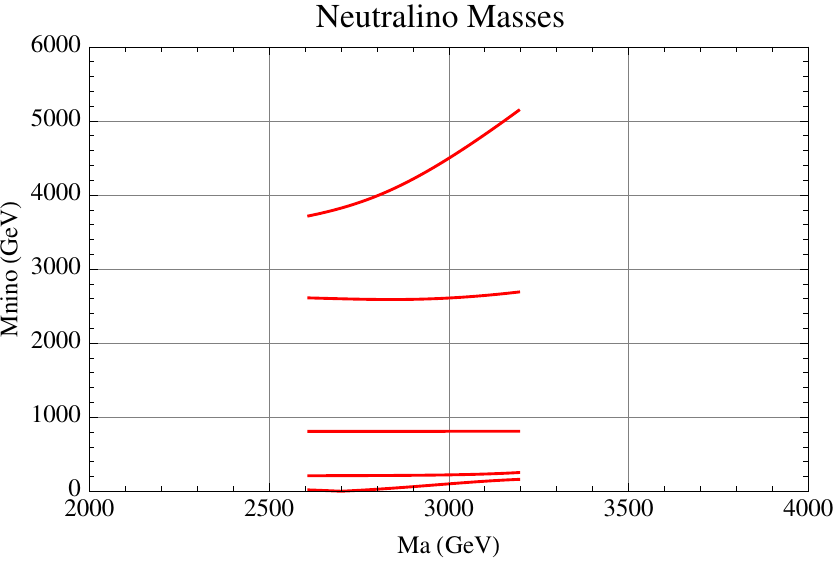} \\
\includegraphics[scale=0.75]{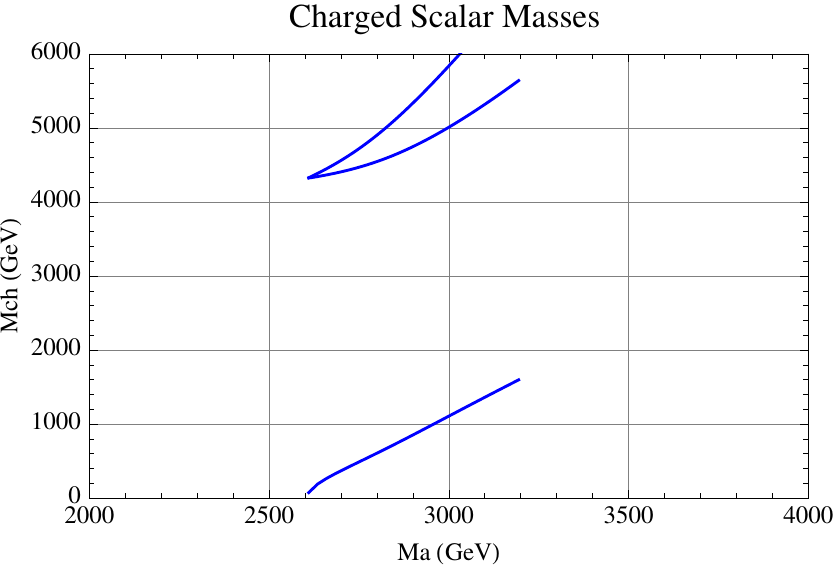} &
\includegraphics[scale=0.75]{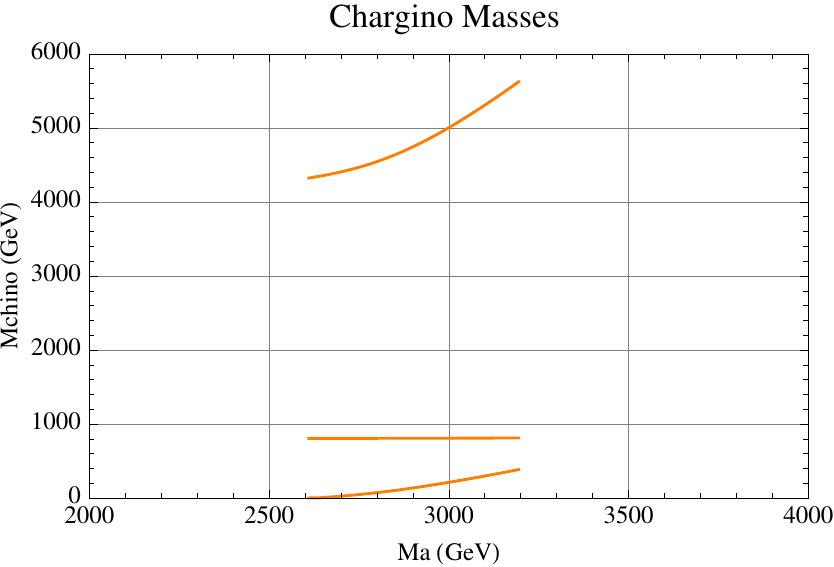} 
\end{array}$
\caption{Masses  as a function of the lightest pseudoscalar mass $m_a$ for a $\mu_{12}$ scan along the yellow line across region D in Fig. \ref{StabilityandLSPRegion}.  The parameters for the plots are $\tan{\beta}=10$, $\tan{\theta}=2$, $b=4,000$ ${\rm GeV^2}$ and $\mu_{11} =-344$ GeV. The  curves correspond to the various particles  just as described in the caption to Fig. \ref{MassesTanBeta1}.}
\label{MassesNegativeb}
\end{center}
\end{figure*}

The model action\cite{CLtV} includes the $SU(2)_L\times U(1)$ SUSY Yang-Mills action, the SUSY nonlinear sigma model constructed in terms of the gauged K\"ahler metric\cite{CLtV}-\cite{Z}, the superpotential and the SUSY breaking terms. After eliminating the auxiliary fields and imposing the constraint, the tree level potential is constructed. There are three minimum condition which need to be satisfied. Their implementation can be used to eliminate the SUSY breaking parameters $m_u, m_d$ and the superpotential mixing parameter $\mu_{21}$. That leaves seven remaining parameters which include the MSSM mixing angle $\tan\beta =v_u/v_d$, the MSSM mu parameter, $\mu_{11}$, and the MSSM SUSY breaking parameters $M_1, M_2$ and $b=-\mu_{11} B$. In addition, there are two new parameters which are the superpotential parameter $\mu_{12}$ and a new mixing angle $\tan\theta =\sqrt{(v_u^2+v_d^2)/2v^{\prime 2}}$. The total VEV is given by $v=\sqrt{v_u^2+v_d^2+2v^{\prime 2}}$ so that $M_Z=\sqrt{g_1^2+g_2^2} v /2= M_W/\cos\theta_W$, where $g_1, g_2$ are the $U(1), SU(2)_L$ gauge couplings. On the other hand, the Yukawa coupling constants through which the quarks and leptons acquire their masses are proportionately enhanced. For example, the top quark Yukawa coupling is $\sqrt{2}m_t/(v\sin\theta\sin\beta)$ which is larger than the standard model or MSSM relation.

The stability region in parameter space is determined by requiring all scalar squared masses to be positive. The model also exhibits an unbroken R-parity which insures the stability of the lightest SUSY partner (LSP). Requiring this LSP to be a neutralino further restricts the allowed parameter space.  Four stability regions, denoted as  A, B, C, and D, are exhibited in Fig.~\ref{StabilityandLSPRegion}. For each panel in the figure, the value of the gaugino SUSY breaking masses are $M_1 =200$ GeV and $M_2=800$ GeV.  Stability region A has $\tan{\beta}=1$, $\tan{\theta}=1$, region B has $\tan{\beta}=1$, $\tan{\theta}=2$, region C has $\tan{\beta}=2$, $\tan{\theta}=2$, and region D has $\tan{\beta}=10$, $\tan{\theta}=2$. Each region is considered for three values of the SUSY breaking parameter $b=-4,000,~4,000,~12,000$ ${\rm GeV^2}$. In general, the eigenvalues of the mass matrices must be determined numerically. Detailed mass spectra for specific points in parameter space indicated by  green dots in Fig. \ref{StabilityandLSPRegion} are displayed in Fig. \ref{MassSpectrum}. Note that the lightest spin zero particle can be either the neutral pseudoscalar $a$ (panels A,B) or the neutral scalar $h$ (panels C, D). The next heaviest neutral pseudoscalar is denoted by $A$, while the remaining neutral scalars in order of increasing mass  are denoted as $H1, H2$. Adapting a similar notation, the neutralinos in order of increasing mass are denoted as ${N1}, {N2}, {N3}, {N4}, {N5}$, while  the charged scalars (charginos) are  $C1, C2, C3$ ($\tilde{C1}, \tilde{C2}, \tilde{C3}$). To further explore the mass spectra, the neutral (pseudo-) scalar, charged scalar, neutralino, and chargino masses as a function of the lightest pseudoscalar mass are exhibited in  Figs. \ref{MassesTanBeta1} -- \ref{MassesNegativeb}. The various curves in the figures follow the parameter scans from left to right for fixed $\mu_{11}$ with increasing $\mu_{12}$ over the range  indicated by the yellow lines in  Fig.~\ref{StabilityandLSPRegion} for each of the four regions A, B, C, and D. The left endpoint of all the curves in each of the figures is dictated by the stability bounds as is the right endpoint of the curves in Figs. \ref{MassesTanBeta10}-\ref{MassesNegativeb}. On the other hand, the right endpoints of the curves in Fig. \ref{MassesTanBeta2} correspond to the maximum value for $\mu_{12}$ plotted in Fig. \ref{StabilityandLSPRegion}. 
\begin{figure*}
\begin{center}
$\begin{array}{cc}
\includegraphics[scale=0.75]{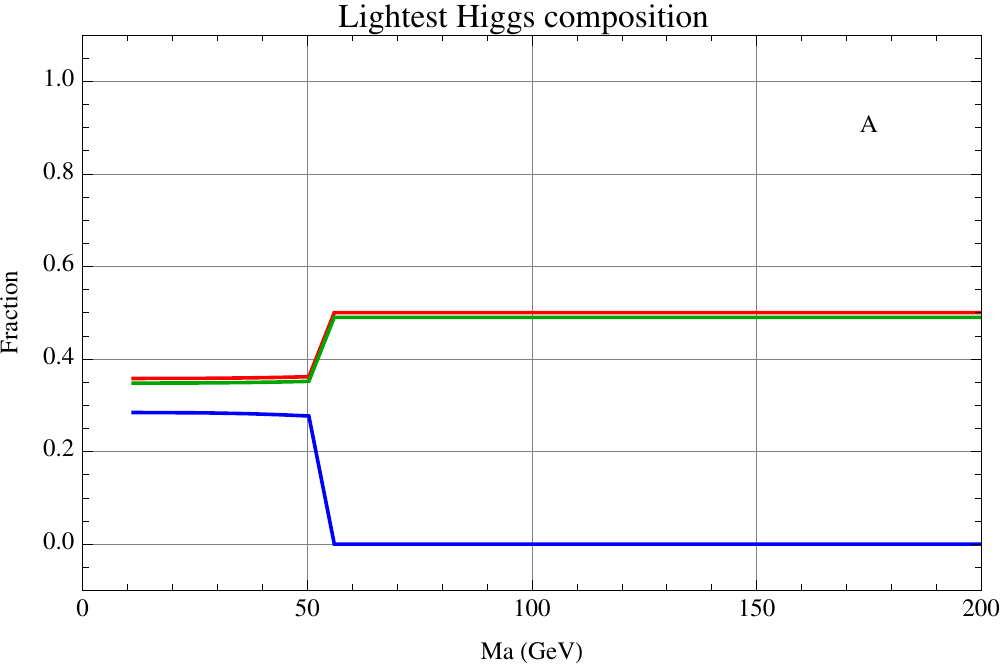} &
\includegraphics[scale=0.75]{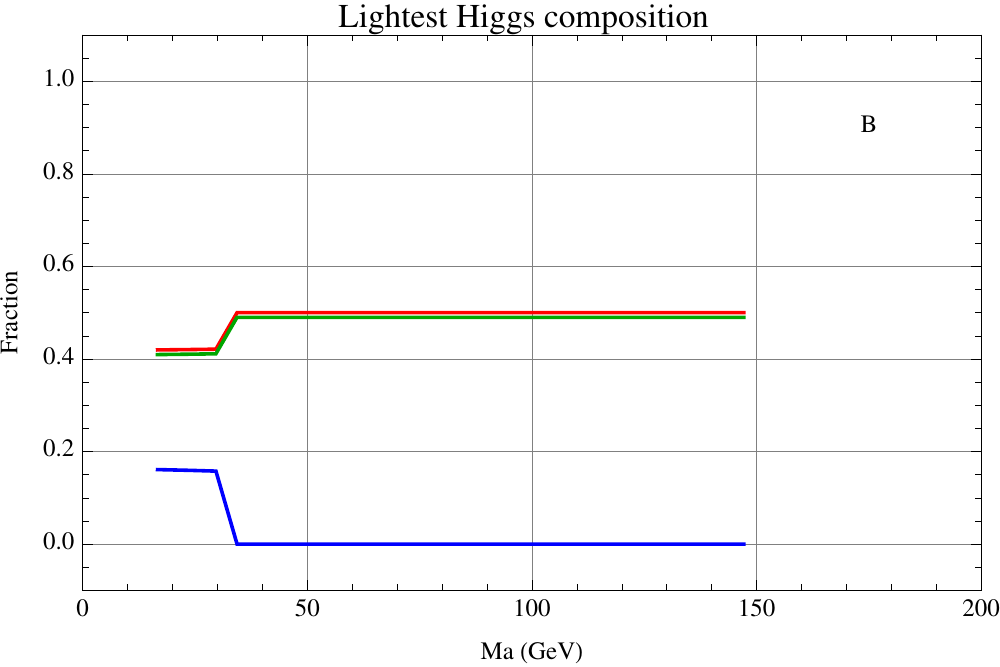} \\
\includegraphics[scale=0.75]{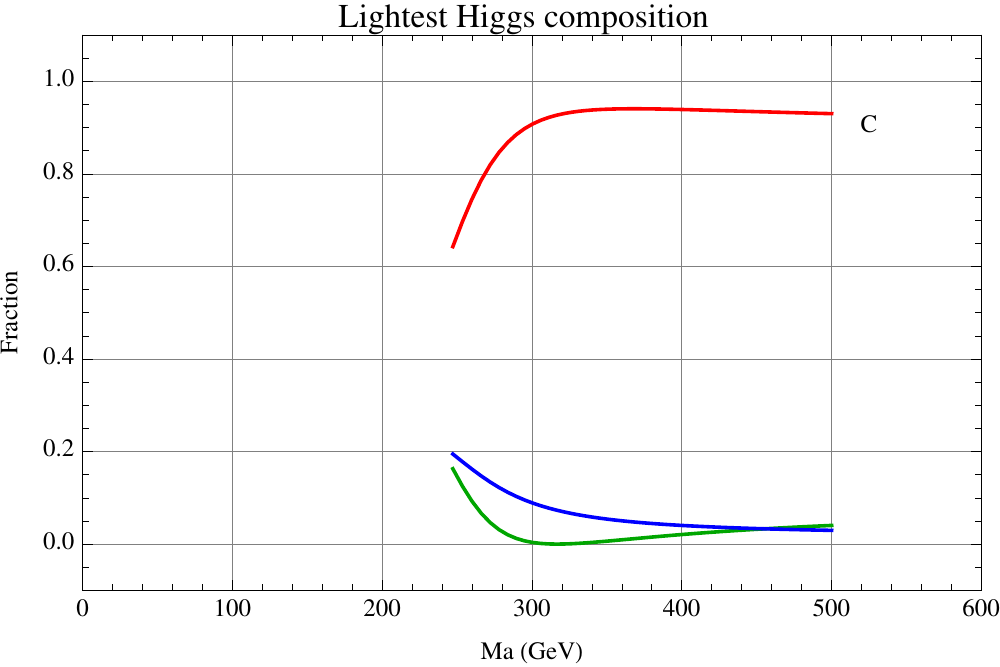} &
\includegraphics[scale=0.75]{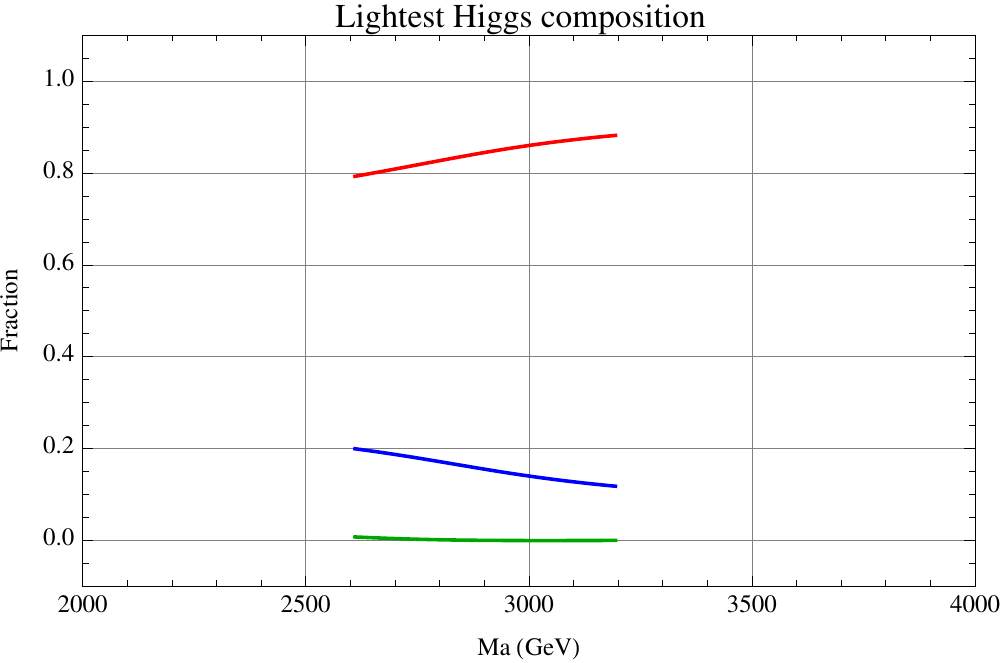} 
\end{array}$
\caption{Lightest neutral Higgs boson, $h$, content as a function of the lightest pseudoscalar mass for a $\mu_{12}$ scan corresponding to the yellow lines across the A, B, C, and D regions in Fig. \ref{StabilityandLSPRegion}. For each plot the values of the gaugino SUSY breaking masses are $M_1 =200$ GeV and $M_2=800$ GeV, and $b=4,000$ ${\rm GeV^2}$.  The scan through region A has $\tan{\beta}=1$, $\tan{\theta}=1$, and $\mu_{11}=-12$ GeV, the one through region B has $\tan{\beta}=1$, $\tan{\theta}=2$ and $\mu_{11}=-16$ GeV, the one through region C has $\tan{\beta}=2$, $\tan{\theta}=2$, and $\mu_{11}=-52$  GeV, and the one through region D has $\tan{\beta}=10$, $\tan{\theta}=2$, and $\mu_{11}=-344$ GeV. The red curve corresponds to the $S_u$ fraction, the green curve to the $S_d$ fraction, and the blue curve to the $S_\pi$ fraction. }
\label{HiggsContent}
\end{center}
\end{figure*}
All four panels allow for a lightest Higgs boson, $h$, with mass greater than $130$ GeV which is the MSSM upper bound even after the inclusion of radiative corrections. Using the experimental bound\cite{PDG} on the lightest MSSM pseudo-scalar of $m_a > 93.4$ GeV as the bound for the current model, we see that  region $A$ allows a lightest Higgs boson tree level mass in the range $130 ~{\rm GeV}~ <m_h< 200$ GeV which corresponds to the range $93.4 ~{\rm GeV} ~<m_a < 180$ GeV, while for region $B$, the lightest Higgs boson mass varies from  $130 ~{\rm GeV}~ <m_h< 172$ GeV as $m_a$ ranges from  $93.4 ~{\rm GeV} ~<m_a < 148$ GeV over the scanned region.
\begin{figure*}
\begin{center}
$\begin{array}{cc}
\includegraphics[scale=0.60]{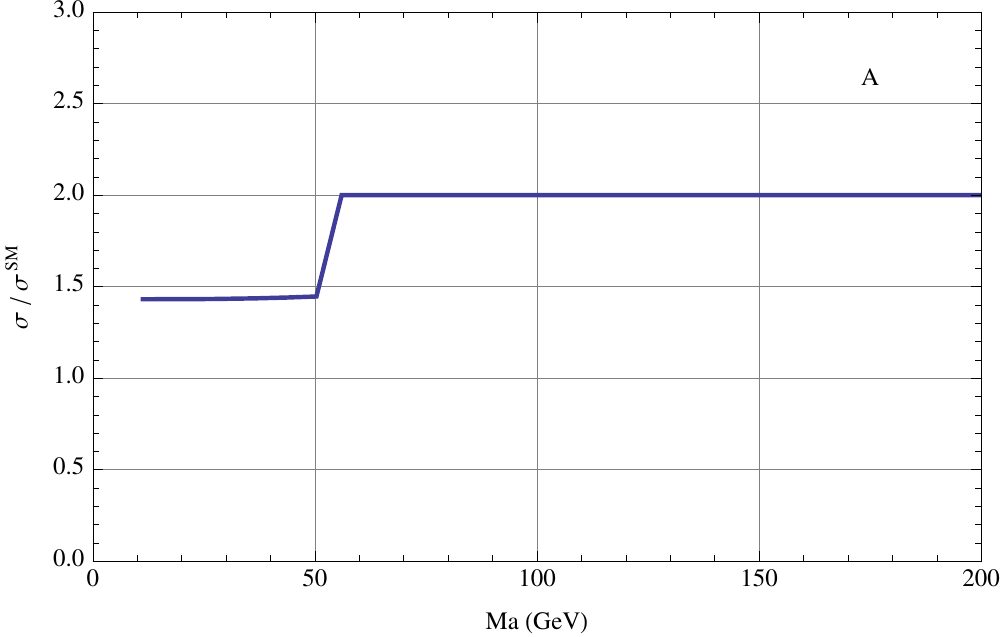} &
\includegraphics[scale=0.60]{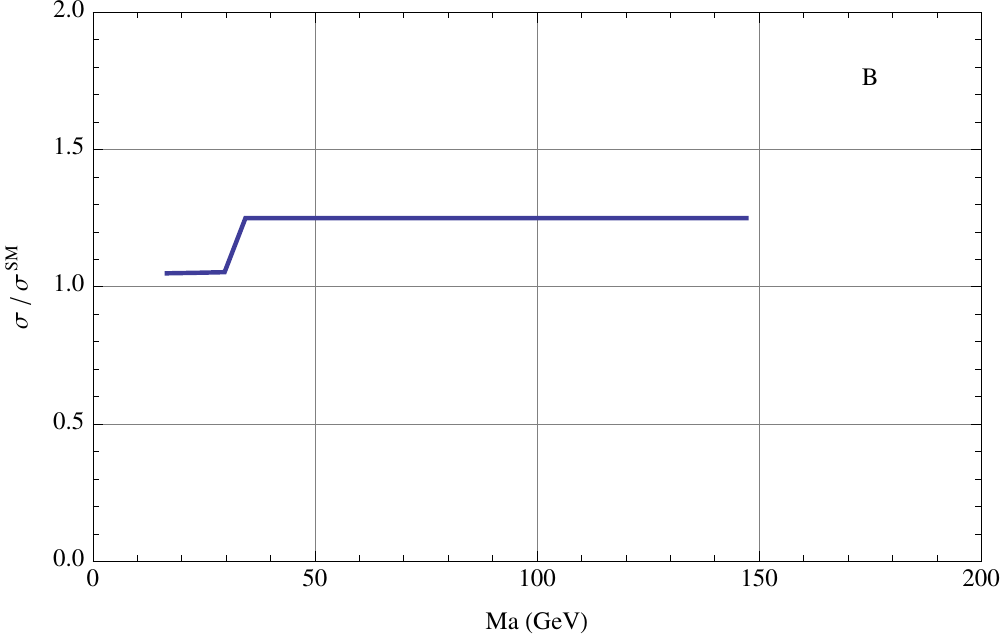} \\
\includegraphics[scale=0.60]{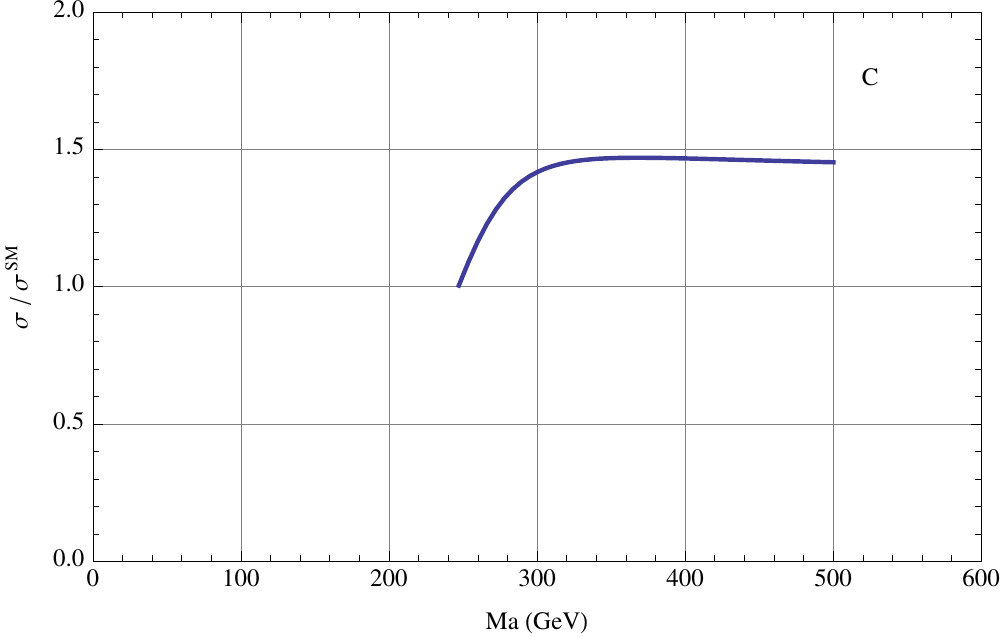} &
\includegraphics[scale=0.60]{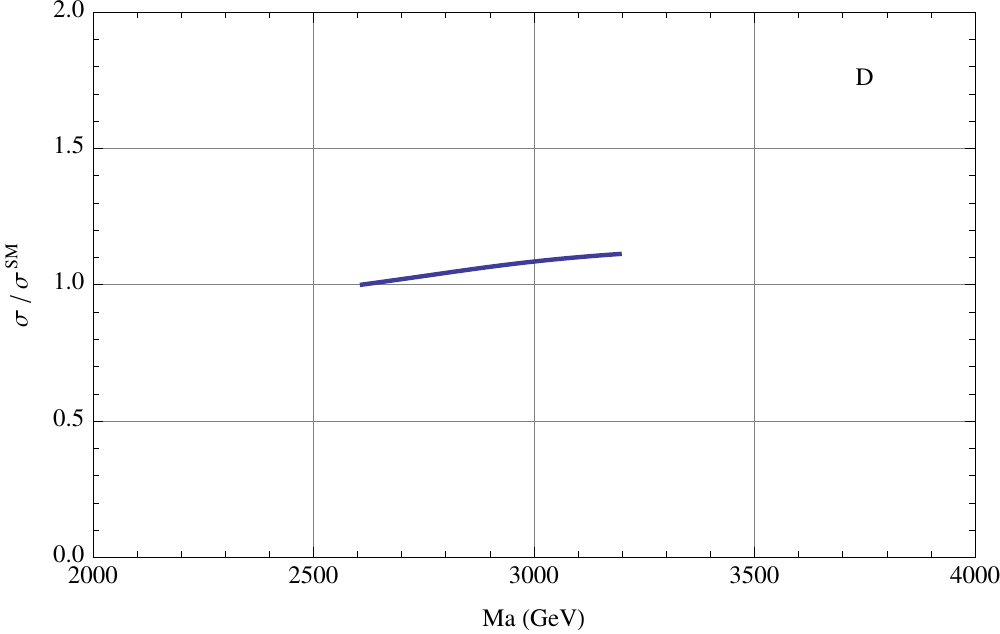} 
\end{array}$
\caption{Ratio of gluon fusion Higgs scalar production cross-section to the standard model result.} 
\label{cross}
\end{center}
\label{fig-cross}
\end{figure*}
A lightest Higgs scalar with a mass in the range $115 ~{\rm GeV}~ <m_h< 130 $ GeV is also allowed provided  different (SUSY breaking) parameters are employed. For the scans considered, region $C$ admits a lightest Higgs boson mass in a range from $182 ~{\rm GeV}~> m_h > 115$ GeV as $m_a$ varies from $370 ~{\rm GeV}~ < m_a < 475$ GeV. For $m_a$ less than around 350 GeV, there is some conflict with the current experimental limit on the mass of the lightest chargino. Finally region $D$  admits a lightest Higgs boson mass in a range from $200 ~{\rm GeV}~>m_h> 115$ GeV as $m_a$ varies from $3140 ~{\rm GeV}~ < m_a < 3180$ GeV. For $m_a$ less than around 3000 GeV, there is some tension with the current experimental limit on the mass of the lightest chargino and/or neutralino.
\begin{figure*}
\begin{center}
$\begin{array}{cc}
\includegraphics[scale=0.60]{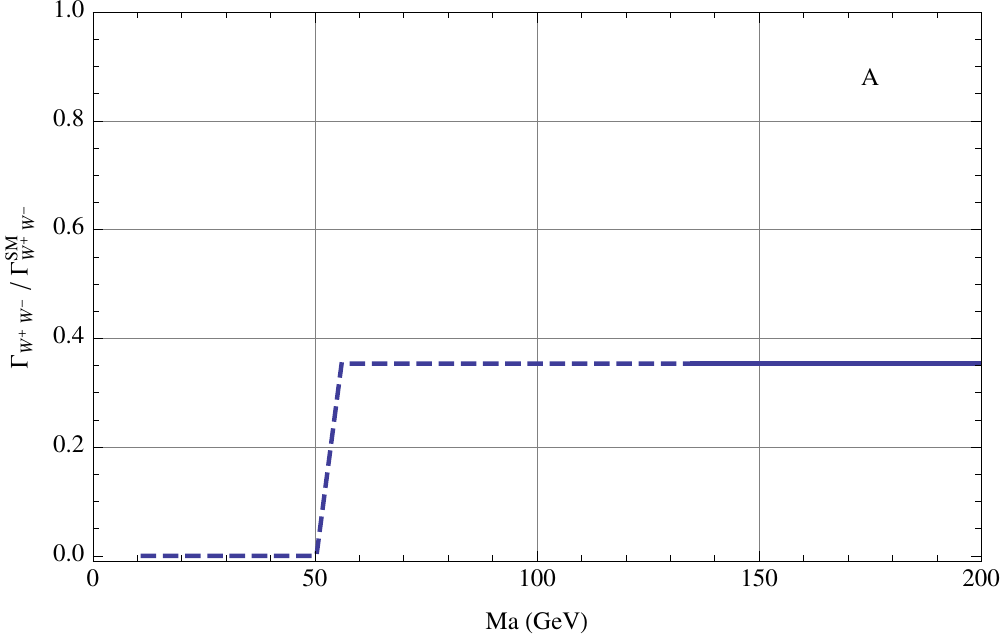} &
\includegraphics[scale=0.60]{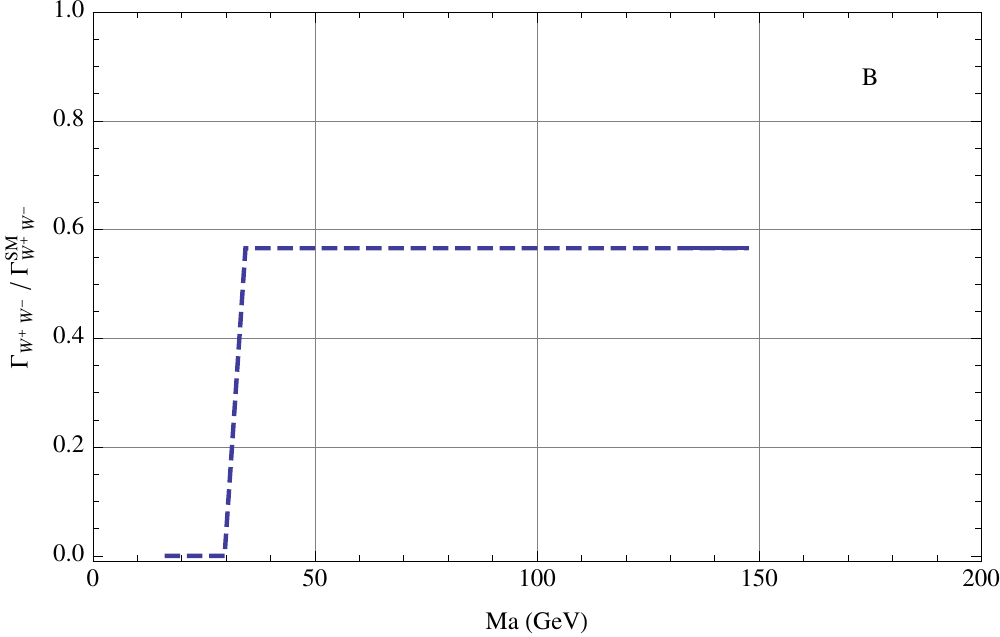} \\
\includegraphics[scale=0.60]{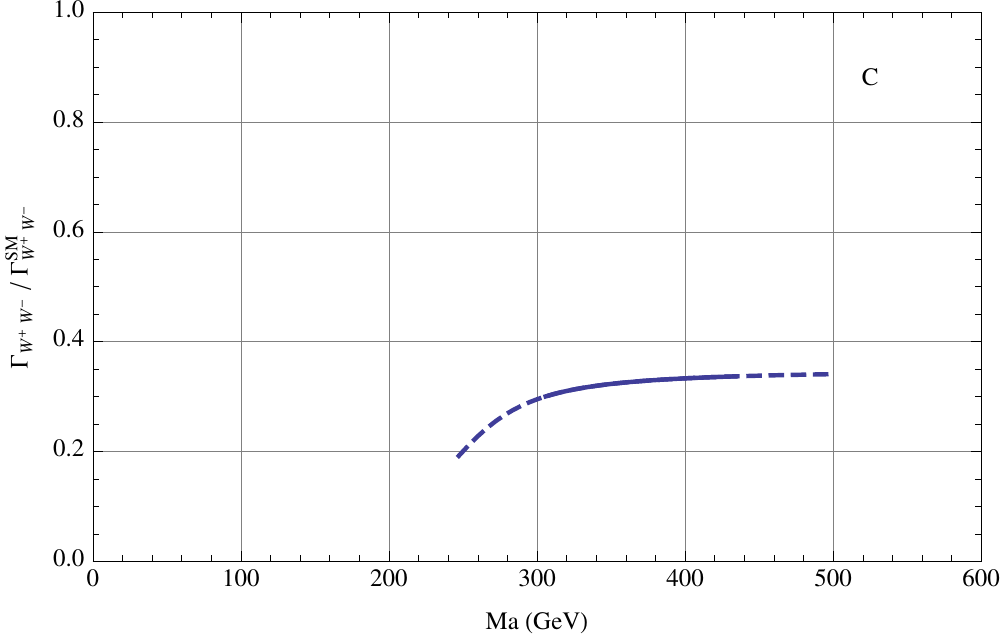} &
\includegraphics[scale=0.60]{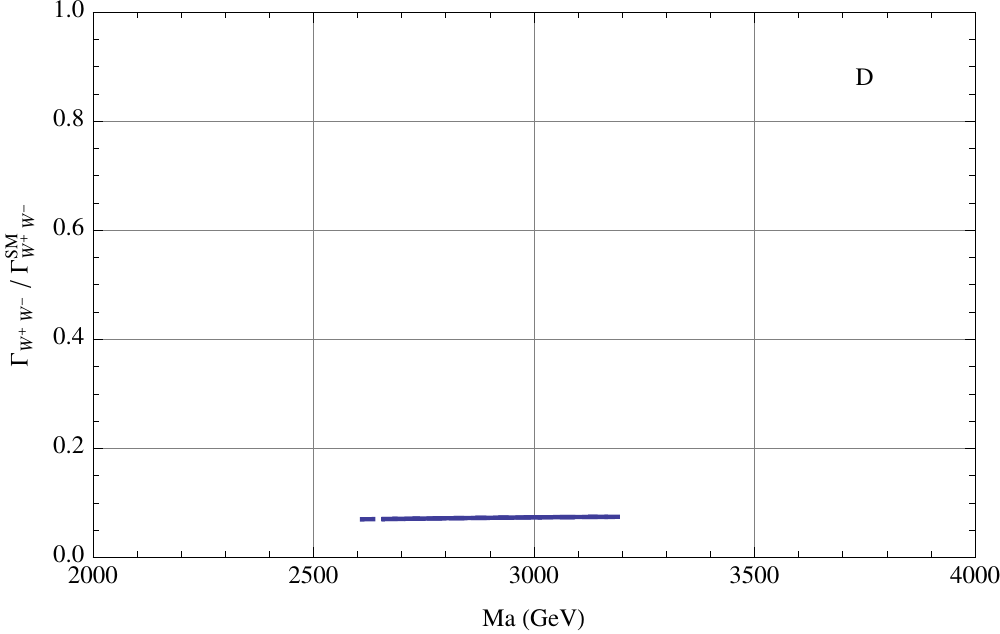} 
\end{array}$
\caption{Ratio of two $W$-boson partial decay width of the Higgs scalar, $h$ to that of the standard model. The dashed line shows the enhancement (suppression) factor over the entire scanned region while solid line corresponds to the  region where the Higgs scalar is sufficiently heavy for the decay to be  kinematically allowed.} 
\label{gammahWW}
\end{center}
\label{fig-gammahWW}
\end{figure*}

\begin{figure*}
\begin{center}
$\begin{array}{cc}
\includegraphics[scale=0.60]{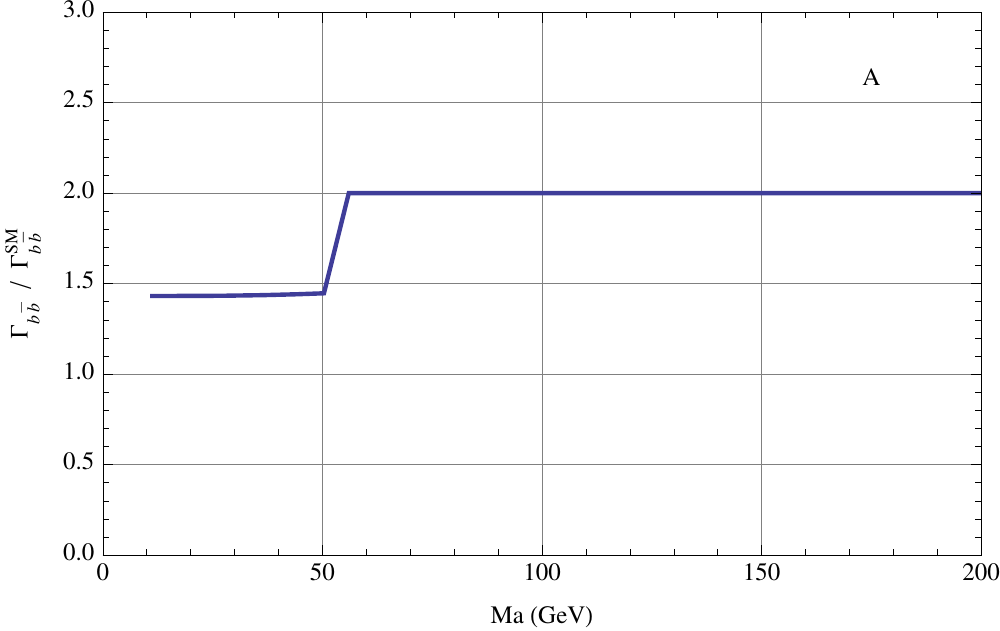} &
\includegraphics[scale=0.60]{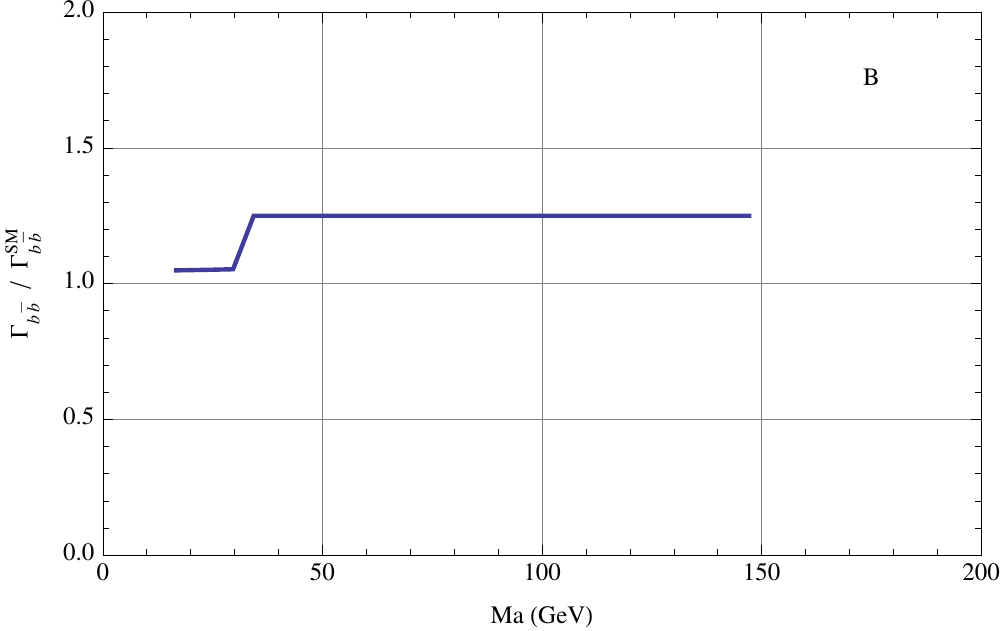} \\
\includegraphics[scale=0.60]{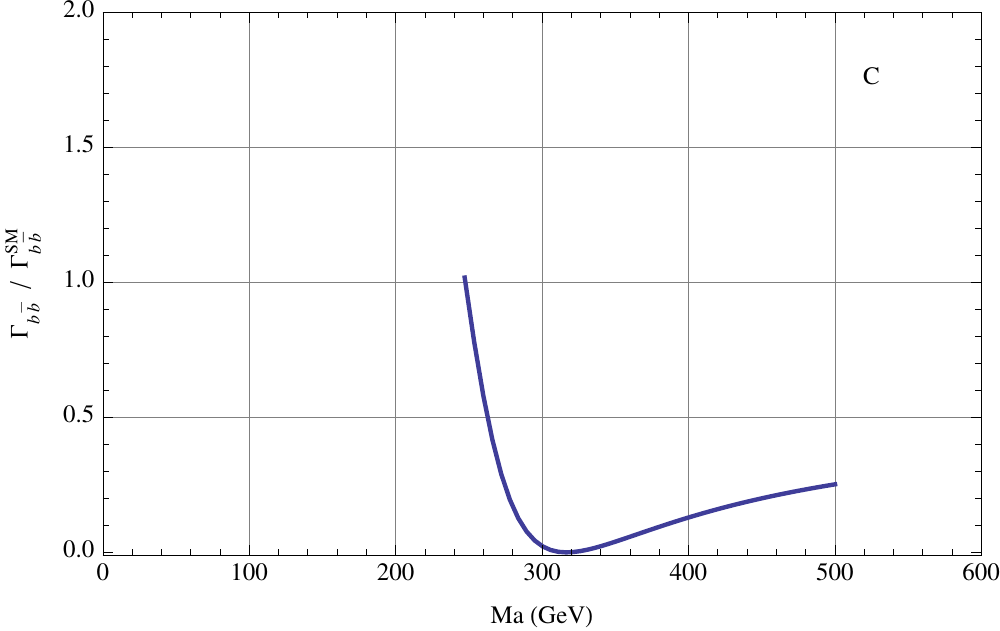} &
\includegraphics[scale=0.60]{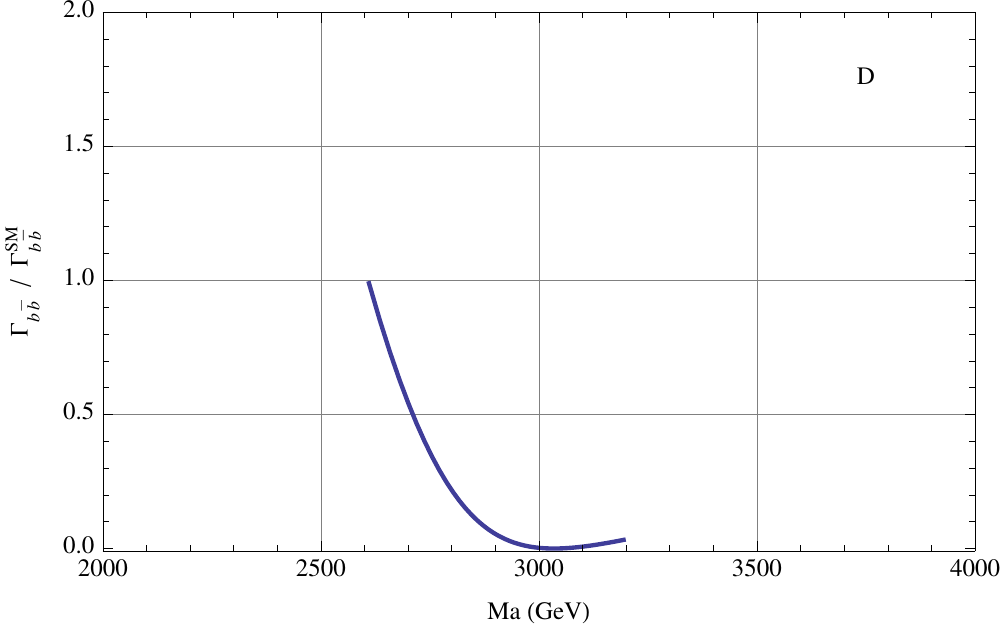} 
\end{array}$
\caption{Ratio of partial width for the decay of the Higgs scalar, $h$,  to two b quarks to that of the standard model.} 
\label{gammavbbbar}
\end{center}
\end{figure*}

It is instructive to quantify the contributions of the components of the MSSM doublets and the constrained Higgs doublet multiplets to the lightest degrees of freedom. Here we focus on the lightest Higgs neutral  scalar which can be decomposed as $h=a_u S_u +a_d S_d +a_\pi S_\pi$ , where $S_u =\frac{1}{\sqrt{2}}(H_u^{0\dagger}+H_u^0), S_d=\frac{1}{\sqrt{2}}(H_d^{0\dagger}+H_d^0)$ and  $S_\pi =\frac{1}{\sqrt{2}}(\pi^{0\dagger}+\pi^0)$. The various fractions $|a_u|^2, |a_d|^2, |a_\pi|^2$  are displayed in Fig. \ref{HiggsContent}  as a function of $m_a$. For regions $A$ and $B$,  a lightest Higgs scalar is essentially devoid of the nonlinearly transforming scalar $S_\pi$ over the entire range  $ 93.4 ~{\rm GeV}~  <m_a$. As such, the composition of the Higgs scalar is thus almost identical to that of the MSSM.  In  region $C$, the $S_\pi$ fraction of is less than  $6-4\%$ for a lightest Higgs scalar mass in the range $182 ~{\rm GeV}~> m_h > 115 $ Gev which corresponds to $370 ~{\rm GeV}~ < m_a < 475$ GeV. 
 While not completely negligible, the Higgs scalar is still predominately composed of the MSSM fields. Finally, for region $D$, the $S_\pi$ content in the lightest Higgs scalar is about $13-12\%$ for the mass range $200 ~{\rm GeV}~ > m_h  > 115 $ GeV which corresponds to $3140 ~{\rm GeV}~ < m_a < 3180$ GeV. The modification to this lightest Higgs production and decay due to the admixture of the non-MSSM content will be addressed sunsequently. The discontinuity in the slope appearing in the plots for regions $A$ and $B$ is a consequence  of the crossover in the particle content of  the lightest mass eigenvalue and the $m_a$ step size used in the numerical calculation. Note that this slope discontinuity occurs at a value of $m_a$ which is less than $93.4$ GeV and hence excluded by the current experimental bound. 
The fractions of the lightest pseudoscalar, charged scalar, lightest neutralino and lightest chargino in terms of the MSSM fields and the constrained fields can be found in reference\cite{CLtV}. The field composition of the neutralino LSP is very similar to that of the MSSM for regions $A, B, C$. Consequently, its  identification  with  dark matter can proceed just as in the MSSM. For region $D$,  the fraction of $\pi-$ino is somewhat larger being of order $10-5\%$ for $3100 ~{\rm GeV}~ < m_a < 3150$ GeV.

Finally, we briefly address the modifications to Higgs boson production and decay.  For moderate $\tan{\beta}$ values, the top quark loop gives the dominant contribution to gluon fusion Higgs production at the LHC provided the squark masses are sufficiently high \cite{Spira:1995rr}.  The lightest Higgs boson can be written as a linear combination of the MSSM scalars $S_u$, $S_d$ and nonlinearly transforming scalar $S_\pi$  as  $h = a_u S_u + a_d S_d + a_\pi S_\pi$. 
The modulus squares of various amplitudes are presented in Fig. \ref{HiggsContent} for the four regions of parameter space numerically probed in this paper. 
Since the top quark interacts only with the $S_u$ component with the enhanced Yukawa coupling $\sqrt{2}m_t/(v\sin{\theta}\sin{\beta})$, the tree level gluon fusion production cross section is equal to that of the standard model times an overall factor so that $\sigma =  |a_u|^2 \left(1 + \frac{1}{\tan^2{\theta}}  \right)\left(1 + \frac{1}{\tan^2{\beta}}  \right) \sigma^{\rm SM}$.
Note that the production rate depends on the details of the MSSM Higgs scalar $S_u$ content for the chosen values of parameter space.  It is clear from Fig.~\ref{HiggsContent} that since $S_u$ comprises at least one-half the Higgs scalar, there will be an enhanced gluon fusion  production rate relative to the standard model as seen in Fig.~\ref{cross}. Modifications to other Higgs production processes such as Higgsstrahlung off a vector boson or top quark, or in the decay of a heavy charged Higgs boson, can also be considered. When considering the decay of the Higgs scalar, $h$, differences from the standard model  arise from  both the presence of the two mixing angles, $\beta, \theta$, in the vacuum expectation values as well as the various particle content of $h$ mentioned above. Since  $v_u^\prime=v_d^\prime= v^\prime$, the coupling of $S_\pi$ to the $W^+ W^-$ pair identically cancels. Consequently, the process $h \rightarrow W^+ W^-$ proceeds only through the  $S_u$ and $S_d$ field components and the tree level decay rate of a heavy Higgs boson to $W^+ W^-$ is the standard model rate modified by a suppression factor so that $\Gamma_{W^+ W^-} = \left(\frac{\tan^2{\theta}}{1+\tan^2{\theta}}\right)\left(\frac{1}{1+\tan^2{\beta}}\right)\left| a_u  \tan{\beta} + a_d ~\right|^2 \Gamma_{W^+ W^- }^{\rm SM} $ as seen in Fig. \ref{gammahWW}. 
Likewise, the decay to $\bar{b}b$ quarks also depends on the $b$-Yukawa coupling, $\sqrt{2}m_b/(v\sin\theta\cos\beta)$, enhancement and the constituent fraction of the $S_d$ content of the Higgs field.  This leads to the modified tree level rate given by
$\Gamma_{b\bar{b}} = |a_d|^2 \left(1 + \frac{1}{\tan^2{\theta}}  \right)\left(1 + {\tan^2{\beta}} \right) \Gamma_{b\bar{b}}^{\rm SM}$  
as displayed in Fig. \ref{gammavbbbar} using the parameter scans appropriate to the four regions. For regions $A$ and $B$, the $b$-pair partial rate is enhanced relative to that of the standard model, while for regions $C$ and $D$, the rate is suppressed. This suppression is a consequence of the very small admixture of $S_d$ in $h$ for these regions.

A model consisting of a supersymmetric nonlinear sigma model incorporating the low energy effects of an unspecified electroweak symmetry breaking sector and coupled to a supersymmetric version of the standard model was constructed and analyzed. The superpotential coupling of the constrained pair of Higgs doublets to the  MSSM Higgs doublet pair catalyzes nontrivial vacuum expectation values in the later thus producing an additional contribution to the electroweak symmetry breaking and also serving as the source of the MSSM matter fields masses.  The tree level particle spectrum of the model was obtained for a variety of model parameters. The MSSM upper limit on the mass of the lightest Higgs scalar  was obviated. Throughout the region of the explored parameter space, the lightest Higgs scalar and the neutralino LSP, which can be identified as a dark matter candidate, were primarily composed of the MSSM fields with only a small admixture of the nonlinear transforming components.  The main difference from the standard model predictions in both Higgs boson production from gluon fusion and Higgs scalar decay to either $W^+W^-$ or $\bar{b}b$ resulted from the constituent nature of the Higgs scalar and the variant Yukawa couplings. Depending on the process and region of parameter space, these differences could lead to either an enhancement or a suppression.

\begin{acknowledgments}
 This work was supported in part by the U.S. Department of Energy under grant DE-FG02-91ER40681 (Theory and Phenomenology Task). I thank T.E. Clark  and T. ter Veldhuis for an enjoyable collaboration.
\end{acknowledgments}

\end{document}